\documentclass[twocolumn]{aa}
\usepackage{epsfig}
\usepackage{graphicx}

\begin{document}
%-----------------------------------------------------------------------
%
%           our macros
%
  \newcommand{\block}{\vrule width 0.5 true cm height 6pt depth 0pt \ }
  \newcommand{\Msun} {$M_\odot$}
  \newcommand{\msun} {M_\odot}
  \newcommand{\Mhi} {$M_{\rm HI}$}
  \newcommand{\mhi} {M_{\rm HI}}
  \newcommand{\Oh} {[O/H]}
  \newcommand{\oh} {{\rm [O/H]}}

  \def\HI{H{\sc i} }
  \def\HII{H{\sc ii} }
  \def\Ha{H$_{\alpha}$ }

  \title{Effects of episodic gas infall on the chemical abundances in
     galaxies}

  \author{J. K\"oppen
          \inst{1,2,4}
          \and 
          G. Hensler
          \inst{3,4}
  }

  \institute{Observatoire Astronomique de Strasbourg, 
            11 Rue de l'Universit\'e, 
            F--67000 Strasbourg, France\\
             email: koppen@astro.u-strasbg.fr
         \and
            International Space University, 
            Parc d'Innovation, 
            1 Rue Jean-Dominique Cassini, 
            F--67400 Illkirch-Graffenstaden, France
         \and 
             Institut f\"ur Astronomie,
             Universit\"ats-Sternwarte Wien,
             T\"urkenschanzstr. 17,
             A--1180 Vienna, Austria\\
             email: hensler@astro.univie.ac.at
         \and
             Institut f\"ur Theoretische Physik und Astrophysik, 
             Universit\"at Kiel, 
             D--24098 Kiel, Germany
  } 

  \date{Received October 28, 2004 / Accepted December 23, 2004}

  \abstract{
      The chemical evolution of galaxies that undergo an episode
      of massive and rapid accretion of metal-poor gas is investigated 
      with models using both simplified and detailed nucleosynthesis 
      recipes. The rapid decrease of the oxygen abundance during infall
      is followed by a slower evolution which leads back to the closed-box 
      relation, thus forming a loop in the N/O-O/H diagram. For large
      excursions from the closed-box relation, the mass of the infalling
      material needs to be substantially larger than the gas remaining in 
      the galaxy, and the accretion rate should be larger than the star
      formation rate. We apply this concept to the encounter of high
      velocity clouds with galaxies of various masses, finding that the
      observed properties of these clouds are indeed able to cause 
      substantial effects not only in low mass galaxies, but also in the
      partial volumes in large massive galaxies that would be affected by the 
      collision. Numerical models with detailed nucleosynthesis prescriptions 
      are constructed. We assume star formation timescales and scaled yields
      that depend on the galactic mass, and which are adjusted to reproduce 
      the average relations of gas fraction, oxygen abundance, and effective 
      oxygen yield observed in irregular and spiral galaxies. The resulting 
      excursions in the  N/O-O/H diagram due to a single accretion event 
      involving a high velocity cloud are found to be appreciable, which could 
      thus provide a contribution to the large scatter in the N/O ratio found 
      among irregular galaxies. Nonetheless, the  N/O-O/H diagram remains 
      an important indicator for stellar nucleosynthesis.
    \keywords{ Galaxies: abundances -- 
               Galaxies: evolution -- 
               Galaxies: dwarf Irregular -- 
               Galaxies: ISM
    }
  }

  \titlerunning{Chemistry of Episodic Gas Infall in Galaxies}
  \maketitle

  \section{Introduction}

\begin{figure}
      \centering 
      \includegraphics[angle=-90,width=10cm]{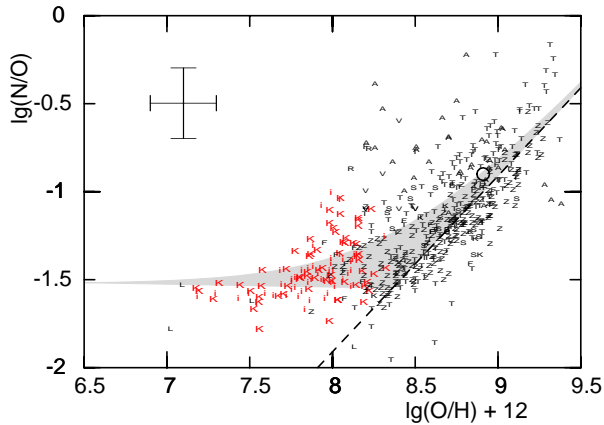}
      \caption[]{The abundance ratio N/O as a function of oxygen abundance
                 observed in spiral galaxies and irregular
                 galaxies after Henry \& Worthey (1999), 
                 but with the symbol for Orion removed 
                 and that for the sun replaced by the large open circle.
                 The dashed straight line is a Simple Model relation for 
                 purely secondary nitrogen production, while the top border  
                 of the shaded area is a relation for a mixture of primary
                 and secondary production.}
      \label{f:hnooho}  
    \end{figure}

  The nitrogen-to-oxygen abundance ratio in extragalactic \HII regions 
  has long been found to increase with the oxygen abundance among spiral
  galaxies (Edmunds \& Pagel 1978, Vila-Costas \& Edmunds 1993,
  Henry \& Worthey 1999 -- Fig. \ref{f:hnooho}).
  Irregular galaxies are located in this plot near the lower
  end of the spiral sequence, and form a cloud with an
  appreciable amount of scatter around $\lg({\rm N/O}) = -1.5$
  and over a range of oxygen abundances between 7 and 8.5.

  Such a behaviour can be interpreted by galactic chemical evolution 
  models as a signature of how the production of nitrogen in stars
  varies with metallicity. The closed-box Simple Model  
  (neglecting the stellar lifetimes) predicts that the relation of
  an abundance ratio $Z_i/Z$ with the abundance of a primary
  element $Z$ mirrors the metallicity dependence of the yield of
  element $i$, strictly independent of the star formation history and the 
  specific recipe for star formation rate (SFR). This also implies that 
  the relation in the diagram can be interpreted as an evolutionary 
  sequence from metal-poor to metal-rich systems.

  Thus, the N/O-O/H diagram has found a good use in infering the mode of 
  nitrogen synthesis in stars. Since the ascending part of the diagram follows
  roughly a unity slope, it is taken to indicate secondary nitrogen
  production (i.e. with a yield proportional to the oxygen abundance).
  The horizontal part, where the irregular galaxies are found, then
  corresponds to the primary nitrogen component, i.e. with constant yield
  (e.g. van Zee 1998). Thus it appears that primary production dominates
  in the early phases of chemical evolution and secondary production in more
  evolved stages.

  This behaviour appears to be somewhat difficult to reconcile with what
  is known from stellar nucleosynthesis studies. While secondary nitrogen 
  production occurs in massive stars, an important contribution comes
  from intermediate-mass stars in the form of a predominantly
  primary production due to the Third Dredge-Up mixing 
  freshly synthesised carbon into higher layers, where it
  can be turned by hot-bottom burning into nitrogen (e.g. van den Hoek
  \& Groenewegen 1997). Thus, once the system is evolved enough for all
  intermediate-mass stars to contribute, the N/O ratio would remain rather
  constant instead of increasing with O/H.

  Several simplifications enter into such an interpretation. One is that the 
  lifetimes of nitrogen-producing intermediate mass stars are not negligibly 
  short compared to the star formation. Such a delayed production makes 
  the N/O ratio rise slower with time (and hence oxygen abundance) until 
  it reaches the Simple Model relation, thus making the slope of the relation 
  somewhat steeper during the early phases (e.g. Matteucci 1986). With this 
  caveat, the N/O-O/H diagram is still an important instrument to probe 
  stellar nucleosynthesis. 

  The other, more important generalization of the Simple Model is the
  accretion of gas, whether by infall from outside the galaxy or outside 
  the disk, or inflow within the disk of the galaxy. K\"oppen \& Edmunds (1999) 
  showed that as long as the ratio of accretion and SFRs decreases 
  monotonically with time -- which occurs in most chemical evolution 
  models, e.g. with the infall rate falling off exponentially with time -- 
  the N/O-O/H relation remains within a factor of two close to the closed-box 
  relation. N/O ratios lower than that are not possible in any kind of infall
  model.

  Detailed chemical evolution models -- which take into account all the
  abovementioned details and processes -- have confirmed that the
  N/O-O/H diagram still remains a very sensitive indicator of stellar  
  nucleosynthesis. Henry et al. (2000) identified the horizontal part at
  $\lg({\rm N/O}) = -1.5$ as the result of primary nitrogen production
  in intermediate mass stars. Because the lag times for nitrogen ejection
  are shorter than the long average star formation timescales  
  in irregular galaxies, the N/O ratio reaches the plateau at lower oxygen
  abundances than in spiral galaxies. The rise of the N/O-ratio in more
  evolved systems is explained by the oxygen yield not being constant, but
  decreasing with metallicity. Such a behaviour would be expected from the 
  evolution of massive stars whose substantial mass loss due to stellar
  winds increases with metallicity (Maeder 1992).

  There remains another potentially important aspect of accretion models: 
  K\"oppen \& Edmunds (1999) also showed that N/O ratios {\rm higher} than 
  obtained from the closed-box relation at the same O/H are possible, if the 
  ratio of accretion and SFR increases so {\it strongly and 
  rapidly} that the metallicity decreases in time. 

  In recent years, it has become more apparent that intergalactic space is 
  pervaded by neutral gas clouds. Among them are the isolated Compact 
  High Velocity Clouds (CHVCs), defined to be kinematically not associated 
  with a galaxy and not to have a stellar counterpart (Braun \& Burton 1999, 
  2000, Kilborn et al. 2000, de Heij et al. 2002). Whether these clouds are 
  left-over building blocks of galaxies (Blitz et al. 1999), remnants of
  galactic winds and fountains, tidal debris, or the product of ram-pressure 
  stripping, their presence implies that they are likely to interact with 
  galaxies. With typical sizes of 15 kpc and gas masses of several 
  $10^7\msun$ with clumps ten times smaller, they could affect
  the chemical evolution of the galaxy. 

  One pertinent feature of the observed N/O-O/H diagram is the large amount 
  of scatter, which is larger than the uncertainties of the observational 
  material and the analysis methods (Henry \& Worthey 1999). In particular, 
  one notes that while there seems to be a plateau at $\lg({\rm N/O}) = -1.5$, 
  there are more objects above this value than below.
  This tendency raises the question whether episodic infall -- the 
  consequence of a collision with an intergalactic gas cloud -- could 
  contribute to the observed distribution of objects in the N/O-O/H plot.

  Over the past years much data on the gas content and the chemical
  composition of irregular galaxies of various masses has become available.
  This material will serve as a constraint to and a comparison with the
  chemical evolution models, in particular the relations between current 
  gas mass and gas fraction, metallicity, and effective yield. 

  In this paper, we first study the basic effects of rapid, massive infall
  on the secondary/primary relation (Section 2) in order to show the basic
  behaviour and to identify the conditions under which strong deviations
  from the closed-box relation can arise. Section 3 presents the systematic 
  relations among the observational data for a sample of gas-rich galaxies.
  Detailed numerical models for the chemical evolution of galaxies which 
  undergo a recent infall episode are described in section 4 along with the
  results and the comparison with observational material.
  Discussion and Conclusions follow as sections 5 and 6.

  \section{Effects of sudden gas infall on abundances}

  The study by K\"oppen \& Edmunds (1999) of the consequences of arbitrary 
  infall of metal-poor gas on the chemical signatures of galaxies showed 
  that the determining factor is the temporal behaviour of the ratio 
  $a(t) = A(t)/\Psi (t) $ of the rates of mass accretion $A$ and star 
  formation $\Psi$. If this quantity decreases monotonically with time, 
  the galaxy evolves with a metal yield smaller than the true yield, but 
  the abundance ratio of secondary and primary elements remains proportional 
  to the primary abundance and is confined to a rather narrow region close 
  to the prediction by the Simple Model 
  $Z_p y_s/y_p \ge  Z_s/Z_p \ge Z_p y_s/(2y_p)$.  

  On the other hand, abundance ratios larger than these limits 
  $Z_s/Z_p > Z_p y_s/y_p$ can only be reached if the accretion is
  massive and rapid enough to cause a decrease of the primary metallicity.
  
  In the following we study more closely the basic behaviour of the chemical 
  evolution of galaxies that undergo such a massive infall of metal-poor gas.
  We consider models in which the finite delay between star formation and 
  metal production is neglected. Oxygen is assumed to be produced primarily, 
  with a yield $y_{\rm O} = 0.0091$, to give results comparable to observed 
  values. Nitrogen is taken to be produced purely in secondary mode with 
  $y_{\rm N} = 0.095 \cdot Z_{\rm O}$. The SFR is supposed
  to be proportional to the gas mass, with a timescale of 5 Gyr. We note that 
  this choice of the SFR does not determine the basic behaviour in the 
  N/O-O/H diagram, although it does affect the exact shape and timing of the 
  evolutionary tracks.

\begin{figure}
      \centering
      \includegraphics[angle=-90,width=10cm]{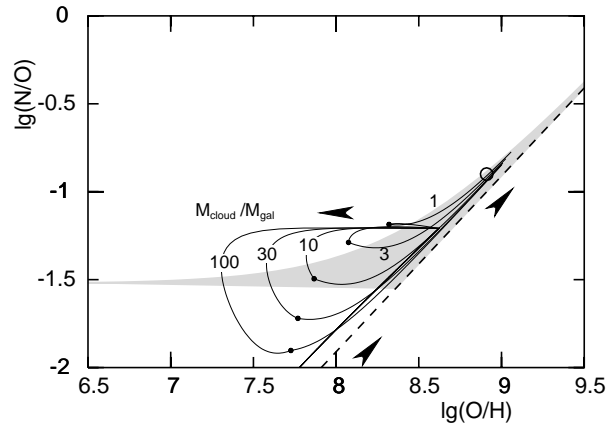} 
      \caption[]{Tracks in the N/O-O/H-diagram of infall models with
          various mass ratios of the infalling gas and the galaxy. 
          The infall event starts at the same age of 5 Gyr and 
          lasts for 1 Gyr (marked by a dot). 
          The shaded area indicates the region of the observations
          as in Fig. \ref{f:hnooho}.}
      \label{f:nolmas}  
   \end{figure}

  In Fig. \ref{f:nolmas} we show the evolution of nitrogen and oxygen
  of models which are closed-box, except for an infall of metal-poor 
  gas starting at 5 Gyr age and lasting for 1 Gyr. It is evident that in 
  order to have a significant deviation from the closed-box secondary-primary 
  relation, the mass of the infalling gas must be substantially larger than 
  the mass of the galaxy. 
  
  Immediately before the infall starts, all models have a gas fraction of 
  0.55 and an oxygen abundance $\oh = -0.3$. The commencement of the 
  infall raises the accretion ratio $a$ from 0 to about 10 and 1000 
  for $M_{\rm cloud}/M_{\rm gal} = 1$ and 100, respectively. We note that
  in galaxies larger in size than a HVC, $M_{\rm gal}$ will represent only 
  that part of the galaxy which mixes with the infalling gas and forms stars.
  The oxygen abundance is strongly reduced due to the `dilution' by the 
  accreted metal-poor gas. As the gas in the galaxy accumulates, the star 
  formation rate increases.
  Since the star formation timescale
  is longer than the duration of the accretion event, the enhancement
  of the SFR takes place appreciably after the infall. Thus we deal not
  with a proper starburst -- in the sense of a reduced star formation
  timescale -- but rather with a galaxy which becomes more gas-rich 
  and thus makes more stars.
  The oxygen synthesized by the massive stars limits the excursion of
  the oxygen abundance, and it also brings down the N/O ratio.
  When the infall stops -- marked by a dot on the tracks -- the gas 
  fraction is between 0.67 and 0.90 for $M_{\rm cloud}/M_{\rm gal} = 1$ 
  and 100, respectively. After this, the models continue to evolve like
  closed boxes, albeit with different initial conditions,
  and the tracks return to the Simple Model secondary-primary
  relationship. The overall effect is that the model traces a `loop' 
  above the Simple Model relation in the N/O-O/H diagram.

\begin{figure}
      \centering
      \includegraphics[angle=-90,width=10cm]{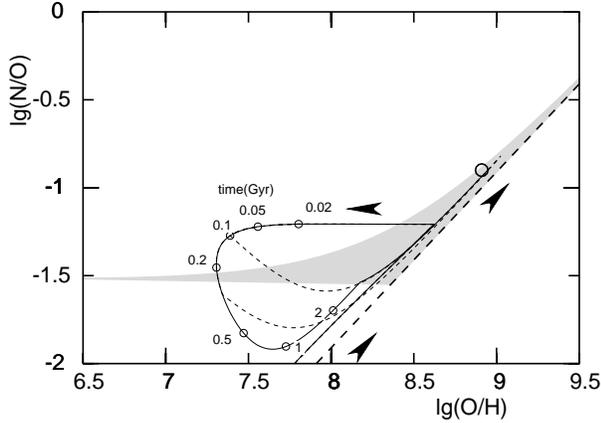} 
      \caption[]{The time evolution of models with an infall rate of 100
          galaxy masses per Gyr, starting at an age of 5 Gyr. The labels 
          refer to the times after the start of the infall.
          The short-dashed curves show the evolution after accretion
          events of 0.1 and 0.3~Gyr.}
      \label{f:noldur}  
    \end{figure}

   For the extreme model with an infall rate of $100\,M_{\rm gal}$/Gyr 
   we follow in Fig. \ref{f:noldur} the evolution along this loop.
   Because of the high infall rate the decrease in \Oh\ 
   takes place rapidly. Depending on the duration of the infall episode, 
   the model returns within less than one star formation timescale to the 
   Simple Model relation. If the infall episode lasts only a short time 
   (e.g. 0.1 Gyr), the increased primary production of oxygen then leads 
   to an evolution towards the lower right (similar to the zig-zag curves of
   Garnett's (1990) starburst model), followed by the (secondary) production 
   of nitrogen leading towards the upper right, towards the Simple
   Model relation. If the infall lasts for a time comparable to the star 
   formation timescale, the oxygen production may just compensate the 
   continuing `dilution' by the infalling gas. The oxygen abundance may 
   remain nearly constant and thus one may obtain a rather large and deep 
   loop leading again towards the close-box relation.

\begin{figure}
      \centering
      \includegraphics[angle=-90,width=10cm]{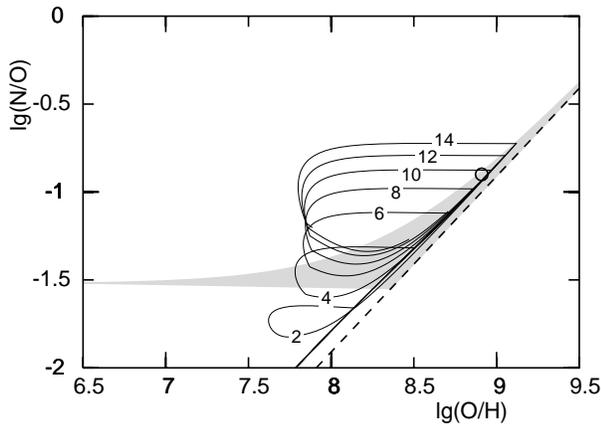} 
      \caption[]{Evolutionary tracks of models with the same mass 
          (10 galaxy masses) and duration of the accretion, but with 
          different starting times (in Gyr, as labeled). All models
          are evolved up to 15 Gyr age.}
      \label{f:noltim}  
    \end{figure}

   The horizontal excursion of the loop depends essentially on the
   mass ratio of infall gas and the gas present in the galaxy. In
   Fig. \ref{f:noltim} we study the influence of the time at which the
   accretion starts. As the gas fraction decreases from 0.80 (at 2 Gyr),
   0.36 (at 8 Gyr) to 0.16 (at 14 Gyr), the loop makes a larger excursion 
   when less gas remains available in the galaxy.

   Finally, we present the evolution for the same model as in  
   Fig. \ref{f:noldur} but with the nucleosynthesis prescriptions we shall 
   use in the further modeling (Section \ref{s:fullmod}): the production of 
   nitrogen is taken from the stellar yields of van den Hoek \& Groenewegen
   (1997). The SFR is assumed to be directly proportional to the gas mass 
   with a timescale of 5 Gyr. Because the primary production from intermediate
   mass stars is important for nitrogen, the loops are less extended 
   in the N/O ratio.
    \begin{figure}
      \centering
      \includegraphics[angle=-90,width=10cm]{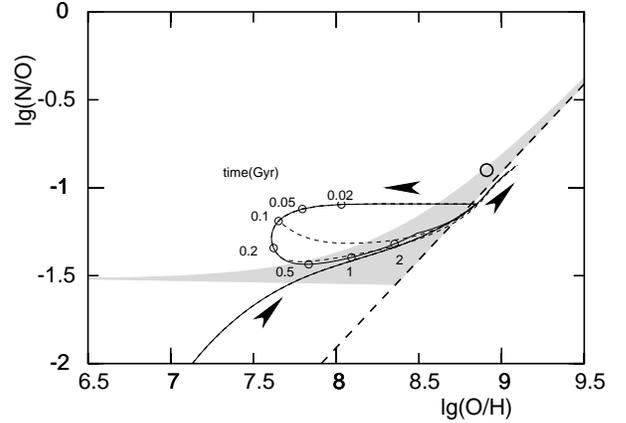} 
      \caption[]{Same as Fig. \ref{f:noldur} but with the proper
                nucleosynthesis recipe we will use in subsequent
                modeling.}
      \label{f:hekdur}  
    \end{figure}
    This figure also demonstrates that the loops done with the full and
    detailed nucleosynthesis prescription obey the characteristics
    predicted from the basic models:
    \begin{itemize}
      \item the N/O ratio does not reach below the limiting locus
            given by the closed-box model relation:
            this region in the diagram cannot be entered by evolution
            (K\"oppen \& Edmunds, 1999).
      \item during the accretion event, the N/O ratio does not
            become larger than the value just before the event.
            The abundance ratio cannot be changed by accretion of 
            metal-poor gas, but only by the production of oxygen
            and subsequently of nitrogen.
      \item the extent of the loop is governed by the mass ratio of
            the accreted gas and the gas left in the galaxy just
            before the event:
            If one neglects metal production during the event,
            infall of $M_{\rm cloud}$ of pure hydrogen gas by a  
            galaxy with a gas mass of $M_1$ and metallicity $Z_1$ 
            leads to a reduced gas metallicity of
            $Z_2 = Z_1 \cdot M_1/(M_1 + M_{\rm cloud})$.
      \item the initial phase of a strong reduction of the
            oxygen abundance is rather short: it is determined by 
            the infall rate which must be high if one wants a
            large excursion from the closed-box relation.
            Hence, it is more likely to observe accreting galaxies 
            during the phase when they evolve back towards the
            closed-box relation.
      \item short episodes of accretion lead to excursions mainly to 
            lower oxygen abundances, while those comparable to the
            star formation timescale are necessary to make excursions 
            towards lower N/O ratio:
            in short infall the gas is only `diluted' whereas during
            longer episodes the resultant secondary (or delayed) production 
            of nitrogen follows only after the self-production of oxygen
            has started.
   \end{itemize}
   We note that genuine starbursts -- periods when the constant
   of the SFR is strongly enhanced -- lead to a different characteristic 
   behaviour, in particular that the N/O-ratio reaches values
   {\it below} the closed-box relation (cf. the zig-zag tracks from
   Garnett, 1990)

  \section{Observed relations among gas-rich galaxies}

  From the literature we collect observational data on gas-rich galaxies. 
  Data on abundances,
  \HI masses, gas fractions etc. are taken from
  Kobulnicky \& Skillman (1996),
  Broeils \& Rhee (1997), van Zee et al. (1997a, 1997b, 2001), 
  Carignan \& Purton (1998), Mateo (1998), Duc et al. (1999, 2001),
  Izotov \& Thuan (1999), van Zee (2001),
  Kennicutt \& Skillman (2001), Garnett (2002), Salzer et al. (2002), 
  Young et al. (2003).
  Data for the same object from different sources are combined, with 
  preference given to the more recent data.

  \subsection{Mass-metallicity relation}

\begin{figure}
      \centering
      \includegraphics[angle=-90,width=10cm]{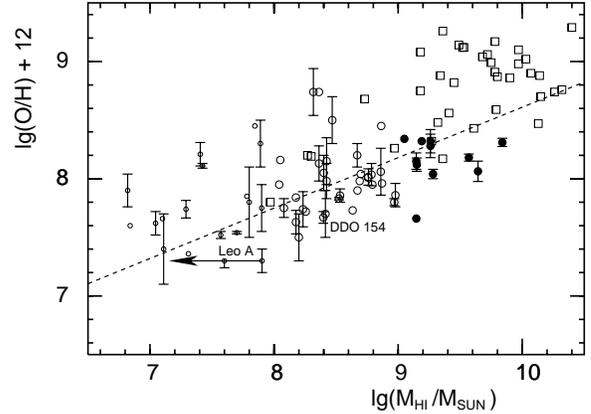} 
      \caption[]{Oxygen abundances and \HI masses of irregular galaxies 
         of our sample. They are divided in three groups according to their 
         \HI mass: small circles ($\mhi < 10^8 \msun$), open circles
         ($10^8 < \mhi < 10^9 \msun$), and filled circles 
         ($\mhi > 10^9 \msun$). Squares denote spiral galaxies.
         Wherever available, individual error bars are given. The dashed line 
         indicates the relation by Garnett (2002).}
      \label{f:mhioho}  
    \end{figure}

   We show in Fig. \ref{f:mhioho} the relation between oxygen abundances
   at the effective radius and the \HI mass of irregular and spiral galaxies. 
   There are 81 galaxies in this sample, and an overall regression
   can be obtained as
   \begin{equation}
     12 + \lg({\rm O/H}) = (4.93\pm 0.37) + (0.38\pm 0.04)\, 
                           \lg({\mhi\over\msun} )
   \end{equation}
   which is almost identical to the relation obtained by Garnett (2002)
   \begin{equation}
     12 + \lg({\rm O/H}) = 4.31 + 0.43\, \lg(\mhi/\msun ).
   \end{equation}
   It is worth noting that spiral galaxies tend to have higher metallicities
   than this relation predicts for their masses. On the other hand, irregular
   galaxies with \HI masses larger than $10^9 \msun$ have lower metallicities.

  \subsection{Gas fractions and effective yields}

  Garnett (2002) deduced gas fractions for 36 objects. For another 
  54 galaxies, we compute them from the ratio of \HI mass to blue 
  luminosity and the extinction-corrected B--V colour index, using 
  the recipe of van Zee (2001):
  \begin{equation}
     {M_{\rm gas}\over M_{\rm stars}} = {\mhi\over L_{\rm B}}
        \cdot 1.3 \cdot {1\over 1.5} \cdot 
        10^{2.84 \, {\rm (B-V)_0} - 1.26}
  \end{equation}
  where the last two terms are based on the photometric properties
  of synthetic stellar populations. 
  The gas fractions are clearly correlated with the \HI masses
  (Fig. \ref{f:mhifgas}), despite a large amount of scatter.
  It might be that this scatter is greater among dwarf galaxies.
    \begin{figure}
      \centering
      \includegraphics[angle=-90,width=10cm]{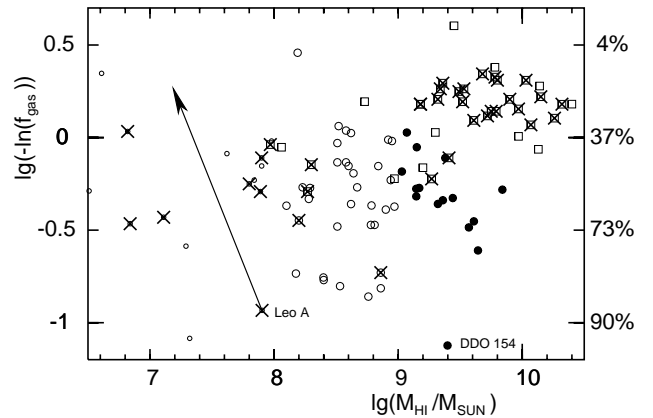} 
      \caption[]{The gas fraction as a function of \HI mass in
                 the galaxies in our sample. Symbols are the
                 same as in Fig. \ref{f:mhioho}. Objects with gas
                 fractions from Garnett (2002) are marked with a 
                 cross}
      \label{f:mhifgaso}  
    \end{figure}
  This behaviour is evident both from the galaxies with gas fractions 
  from Garnett (2002) -- marked by a cross --   as well as 
  in those whose gas fraction we computed from their \HI masses 
  and blue luminosities.

\begin{figure}
      \centering
      \includegraphics[angle=-90,width=10cm]{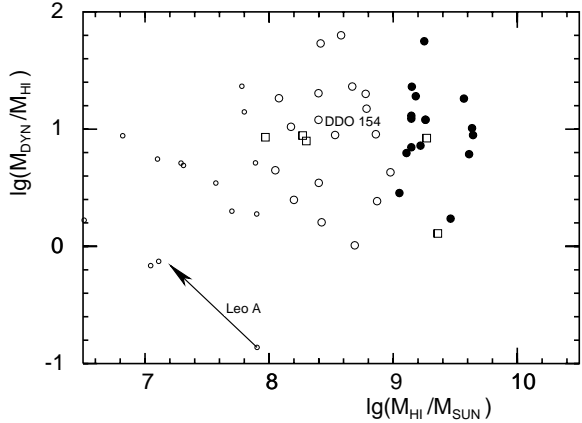} 
      \caption[]{The ratio of dynamical and \HI mass as a 
               function of \HI mass. Symbols are 
               as in Fig. \ref{f:mhioho}.}
      \label{f:mhimrat}  
    \end{figure}

   The \HI component in spiral and irregular galaxies is usually much
   larger in extent than the stellar disk. Thus the observed \HI mass 
   may not be representative of the gas fraction in the galaxy. An
   example is Leo A where Young \& Lo (1996) find that emission 
   from the inner galaxy is about 20 percent of the whole emission. 
   We show its position in all figures for the entire \HI mass, but
   with an arrow pointing towards the loci if we took an \HI mass five 
   times smaller. Fig. \ref{f:mhimrat} shows that this galaxy also has 
   a very low ratio of dynamical to \HI mass, significantly smaller 
   than most other galaxies. However, if one reduces the true galactic gas 
   mass, this object becomes more similar to the others.

\begin{figure}
      \centering
      \includegraphics[angle=-90,width=10cm]{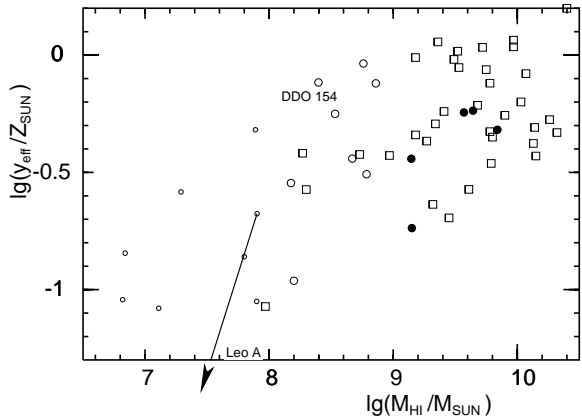} 
      \caption[]{The effective oxygen yield relative to the solar oxygen
                abundance ($Z_{{\rm O},\odot} = 0.00959$) as a function of 
                the \HI mass of the galaxies in our sample. Symbols are 
                as in Fig. \ref{f:mhioho}.}
      \label{f:mhiyeffo}  
    \end{figure}

   With the knowledge of the gas fraction we can compute the effective
   yields, i.e. we compare the oxygen abundances and gas fractions with the
   Simple Model relation.
   Figure \ref{f:mhiyeffo} shows that despite an appreciable amount
   of scatter, there is a clear trend that more massive galaxies have higher 
   effective yields. Spiral galaxies seem to be a natural extension of the 
   irregulars.  
   This relation might indicate that the true yield is smaller in less
   massive galaxies, such as via a steeper IMF, but it could also reflect
   that the effective yield is lower, because the stellar ejecta are less
   likely to be held back by the galaxies' gravitational potential.  

\begin{figure}
      \centering
      \includegraphics[angle=-90,width=10cm]{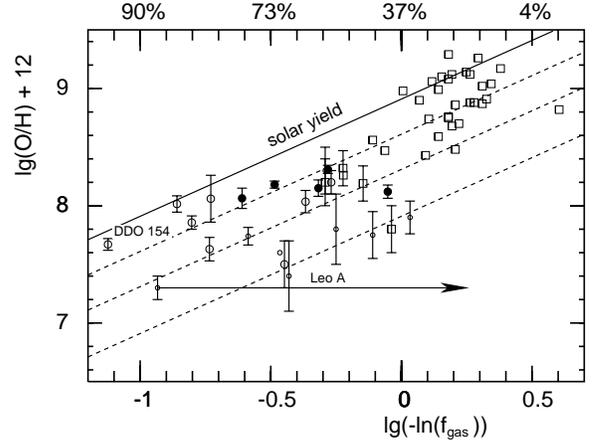} 
      \caption[]{The oxygen abundances and gas fractions 
                (logarithmically on the lower axis, in percentages on the 
                 upper axis) compared to
                 the Simple Model relations for solar yield
                 ($y = Z_{\rm O, \odot} = 0.00959$, solid line) and
                 one half, one quarter, and one tenth solar yields (dashed).
                 Symbols are the same as in Fig.\ \ref{f:mhioho}.}
      \label{f:fgasoho}  
    \end{figure}

   Figure \ref{f:fgasoho} shows the oxygen abundances plotted against the 
   gas fractions, and compared to the Simple Model. It presents 
   essentially the same information as Figs. \ref{f:mhifgaso} and 
   \ref{f:mhiyeffo}, but for a smaller number of galaxies for which all
   three data are available. Among the irregular galaxies one notes a
   clear trend of more massive objects having a higher effective yield,
   but the tendency for them to also have a lower gas fraction is much
   less evident than in Fig. \ref{f:mhifgaso}. Most probably this impression
   is due to the smaller number of objects and the large amount of scatter
   in all diagrams, some of which may well be genuine. The spiral galaxies
   are less gas-rich and have higher effective yields, and they might be 
   interpreted as an extension of the irregular galaxies to higher masses.

\begin{figure}
      \centering 
      \includegraphics[angle=-90,width=10cm]{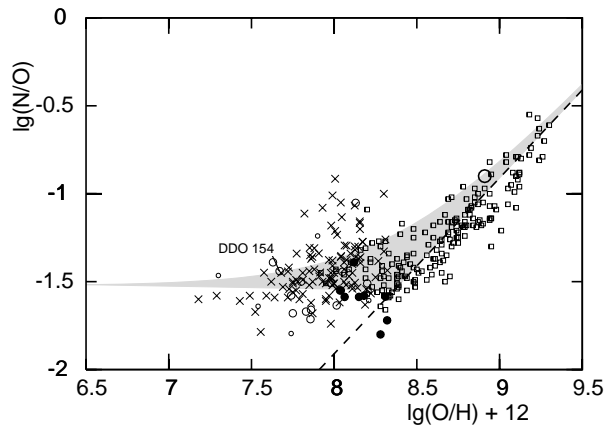}
      \caption[]{Same as Fig. \ref{f:hnooho}, but for the galaxies in the
                 present sample: spiral galaxies (small squares, individual
                 \HII regions from van Zee et al. 1998b) and irregular 
                 galaxies (symbols as in Fig. \ref{f:mhioho}, crosses: 
                 without \Mhi\ determination).}
      \label{f:nooho}  
    \end{figure}

   Finally, we show the N/O-O/H diagram for the galaxies in our sample. For 
   spiral galaxies we show the individual \HII region data of van Zee et al.
   (1998b) in Fig. \ref{f:nooho}.

  \subsection{Diameters}

  Since the excursion from the closed-box relation in the N/O-O/H diagram
  is determined by the mass ratio of infalling material to the gas still
  within the galaxy, we also need to consider what fraction of the galaxy 
  will be affected by the infall of the gas cloud. According to 
  Braun \& Burton (2000), a clump in a high velocity cloud has a typical 
  size of 1~kpc (a few arcmin at 1 Mpc distance). In a collison with a galaxy
  larger than this, only a fraction of the gas in the galaxy will participate
  and thus the chemical evolution is expected to be affected in this partial 
  volume. To estimate this fraction, we need to know the size of the 
  \HI component of the galaxy.
 
\begin{figure}
      \centering
      \includegraphics[angle=-90,width=10cm]{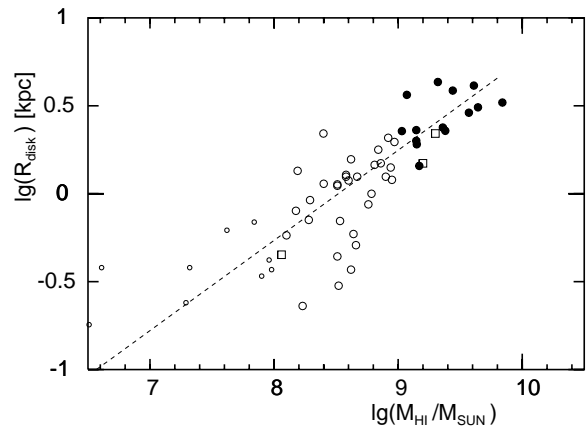} 
      \caption[]{The radial scale length of the stellar disk as a function 
                 of gas mass. The dashed line is the relation of 
                 Broeils \& Rhee (1997) with the assumption of 
                 $R_{\rm disk} =  D_{25} / 6$.}
      \label{f:mhirscl}  
  \end{figure}

\begin{figure}
      \centering
      \includegraphics[angle=-90,width=10cm]{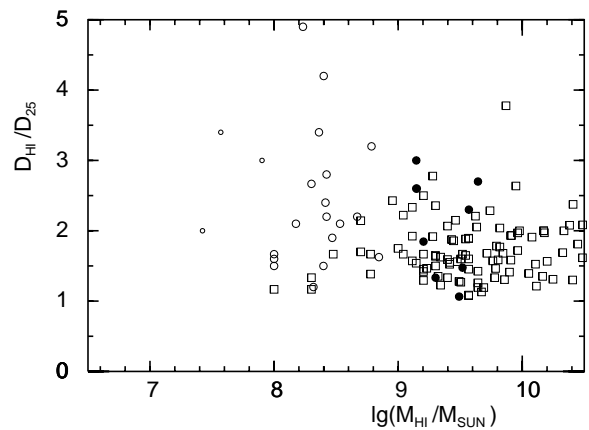} 
      \caption[]{The ratio $D_{\rm HI}/D_{25}$ of the diameter of
                 the \HI distribution and the optical diameter.}
      \label{f:mhidh}  
  \end{figure}

  Broeils \& Rhee (1997) established a tight relation between the diameter and
  the mass of the \HI disk in 108 spiral and irregular galaxies:
  \begin{equation}
     D_{\rm HI} = (\mhi / 10^{6.52} \msun )^{1/1.96}
  \end{equation}
  They also found a tight relation between the diameter of the
  optical disk and the \HI mass:
  \begin{equation}
     \label{e:hidiam}
     D_{25} = (\mhi / 10^{7.00} \msun )^{1/1.95}
  \end{equation} 
  If we assume for the exponential scale length of stellar disk:
  \begin{equation}
     R_{\rm disk} =  D_{25} / 6
  \end{equation}
  we get a very good match with the corresponding data of the objects
  in our sample, as shown in Fig. \ref{f:mhirscl}.
  Most of these data are from van Zee et al. (1997a, 2001) who find 
  that the integrated luminosity profiles 
  of the irregular systems are reasonably well fit by an exponential disk, 
  and determine disk scale lengths by matching the surface brightness 
  distributions in the R-band.

  van Zee et al. (1997b, 1998c) find that the ratio of the diameters seen
  in \HI and the optical $D_{\rm HI}/D_{25}$ is about 2 for `normal' and
  low surface brightness dIrrs and between 3 and 5 for BCDGs. There 
  is no strong trend with \HI mass, as shown in Fig. \ref{f:mhidh}.
  
\section{Models with detailed stellar nucleosynthesis}
  \label{s:fullmod}

   We present calculations of the chemical evolution of galaxies
   of a range of masses, using detailed stellar nucleosynthesis
   prescriptions and follow the evolution due to the accretion of
   a clump of metal-free gas, taking into account that only the 
   star-forming region is affected where the abundances are
   determined from the \HII region, so that in larger
   galaxies only a fraction of the galactic gas is involved.
   These models are constructed under the constraint of matching
   the various observational relations discussed above, and to show
   whether this process explains 
   the scatter among the irregular galaxies in the N/O-O/H diagram.

  \subsection{Assumptions of the underlying model}

   For the prescriptions of stellar nucleosynthesis we follow 
   Henry et al. (2000) and use the results of Maeder (1992) for massive 
   stars and van den Hoek \& Groenewegen (1997)
   for intermediate mass stars. We assume a Salpeter IMF with a stellar
   mass range from 0.1 to 100 \Msun . This gives a ratio 
   $\lg({\rm N/O}) = -1.4$ for the very low mass galaxies, slightly larger 
   than the observed limit of $-1.5$. Although this offset is still well within 
   the observational uncertainties and the accuracy of the nucleosynthesis 
   results, it can be removed by slightly increasing the oxygen yield by a 
   factor of 1.2. Thus the yield for oxygen (at solar metallicity) is 0.026,
   about twice the solar oxygen abundance. Instead of introducing a galactic 
   mass loss of some kind -- with its associated free parameters -- we  
   reduce both oxygen and nitrogen stellar yields by a common factor which 
   depends on the galactic mass $M_{\rm gal, 0}$:
%
%    ##########################################
%
%       notice to printer:
%
%          i had wanted a curved left parenthesis here, but
%          but i always got error messages, if i used 
%                \left{
%          so in the end i gave up fighting latex....
%          and used \left(
%    ##########################################
%
   \begin{equation}
       \label{e:yieldfactor}
       {y \over y_{\rm nucl}} = 
        \left(
         \begin{array}{ll}
              0.1 \cdot M_8^{0.4}  
                           & \mbox{if $M_8<1$}\\
              0.1 \cdot M_8^{0.1}  
                           & \mbox{if $1< M_8< 31.6$}\\
              0.0355 \cdot M_8^{0.4} 
                           & \mbox{otherwise}
          \end{array}
          \right.
   \end{equation}
   with $M_8 = M_{\rm gal, 0}/10^{8} \msun$. The SFR 
   \begin{equation}
       \Psi(t) = {1\over \tau_{\rm SFR}} \cdot (g(t)+s(t)) 
             \cdot \left({g(t)\over g(t)+s(t)}\right)^2
   \end{equation}
   is assumed to depend on the instantaneous masses of gas $g$ and 
   stars $s$, thus on the total mass and the gas fraction. One advantage 
   of this form is a well defined star formation timescale $\tau_{\rm SFR}$.
   In order to give a reasonable match of the observed dependence of gas 
   fraction with gas mass, (Fig. \ref{f:mhifgaso}) we assume this relation
%
%    ##########################################
%
%       notice to printer:
%
%          i had wanted a curved left parenthesis here, but
%          but i always got error messages, if i used 
%                \left{
%          so in the end i gave up fighting latex....
%          and used \left(
%    ##########################################
%
   \begin{equation}
       \label{e:sfrscale}
       \tau_{\rm SFR} =  20\,{\rm Gyr} \cdot 
        \left(
         \begin{array}{ll}
             M_8^{-0.1} & \mbox{if $M_8<1$}\\
             M_8^{-0.4} & \mbox{otherwise}\\
          \end{array}
          \right.
   \end{equation}
   Here we do not assume the factor used by Henry et al. (2000)
   which enhances the SFR for higher metallicities. 
   We also assume that the galaxy is formed by a slow continuous infall 
   with a rate of
   \begin{equation}
       \dot{M}(t) = M_{\rm gal, 0} \cdot {\exp(-t/\tau_{\rm f}) 
                                 \over 1 - \exp(-T/\tau_{\rm f})} 
   \end{equation}
   with a total evolution time of $T = 15$ Gyr and a timescale 
   $\tau_{\rm f} = 4$ Gyr, taken from the solar neighbourhood. This 
   assumption is not crucial for our results: computation with a true 
   closed-box model gives slightly larger oxygen abundances for
   the more massive galaxies.

  \subsection{Modeling the accretion event}

   From Braun \& Burton (2000) we take that a typical CHVC
   has an extent of about 10~kpc (1 degree at 1 Mpc distance) with 
   embedded clumps of about 1~kpc size (a few arcmin to 20'). 
   The clumps have \HI column densities of about 
   $10^{20}$~cm$^{-2}$ and thus masses of about $10^6 \msun$, while the 
   envelope (with $10^{19}$~cm$^{-2}$) is about 100 times more massive.
   For a typical relative velocity of the cloud of 100 km/s, this gives 
   a duration of 10 to 100 Myr for an accretion of the clump and the 
   entire cloud, respectively. It is much faster than the star 
   formation timescale, estimated from the gas consumption, which is 
   several Gyr. This implies that the deviation from the closed-box 
   evolution cannot be a fully developed loop (cf. Figs. \ref{f:noldur} 
   and \ref{f:hekdur}), but rather consists of an excursion to lower 
   oxygen abundances, followed by the track evolving back towards the 
   closed-box relation.

   As a simple approach the clump is assumed to affect only the region 
   within $r = 1$~kpc of the galaxy's centre. For such a central collision, 
   it is straightforward to estimate the mass fraction of the galaxy: 
   \begin{equation}
     \label{e:meff}
     M_{\rm gal, aff}  = M_{\rm gal, 0} \cdot (1 - (1+r/R)\cdot e^{-r/R})
   \end{equation}
   with the scale length $R$ of the exponential \HI disk, which we
   take to be one sixth of the \HI diameter $D_{\rm HI}$. Combining this 
   with the relation (Eq. \ref{e:hidiam}) of Broeils \& Rhee (1997)
   between \HI diameter and mass, we get that while a gas clump would
   affect more than 80 \% of the mass of galaxies below $10^7 \msun$,
   it can influence only 14 \% and 1.8 \% in a galaxy with
   $\mhi = 10^9$ and $10^{10} \msun$. 
   This means that a gas mass of less than $10^8 \msun$ is involved, 
   and hence the accretion of a $10^8 \msun$ gas clump will make a 
   significant impact on the chemical evolution of the galaxy's centre.
   For each galaxy we compute two models: one for the unaffected portion
   is done without an infall event, the other including the infall event.
   The stellar and gas masses of both parts are then added together. 
   The metallicities are taken only from the affected part,
   since this will be what is measured in the current \HII regions.  

   All galaxies evolve following the above presciptions, until the present 
   epoch -- 13 Gyr after the start of the evolution. At that time a 
   metal-poor gas cloud is accreted over a duration of 100 Myrs with a 
   constant rate. The mass of the cloud is randomly picked 
   between $10^6$ and $10^8 \msun$ following a probability distribution 
   proportional to $M_{\rm cloud}^{1.5}$. We do not need to consider smaller 
   masses because of their negligeable effects.
   In all following figures we show the evolution of the models during the
   time interval between 13 and 15 Gyr after start of the evolution, i.e.
   what would be observable from now on.

 \subsection{Results: gas fractions}

\begin{figure}
      \centering
      \includegraphics[angle=-90,width=10cm]{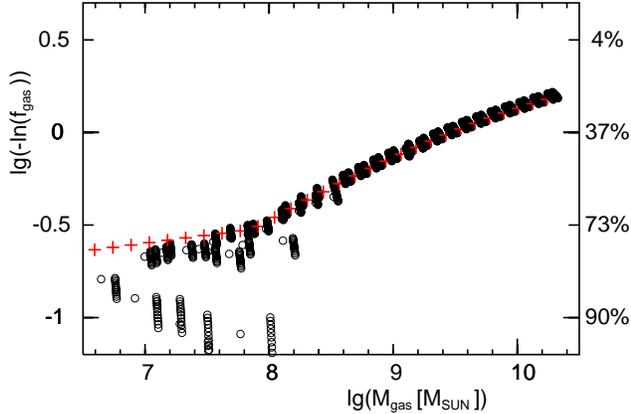} 
      \caption[]{Gas fractions of the models as a function of their gas 
                masses for times between 13 and 15 Gyr after the start of 
                the evolution. The large plus-signs refer to models which 
                do not undergo an accretion episode. The open circles 
                indicate models which are subject to an infall episode at 
                13~Gyr age. The symbols are evenly placed in time over the 
                2~Gyr time span.}
      \label{f:mhifgas}  
    \end{figure}

   Comparison of the gas fractions (Figs. \ref{f:mhifgaso} and 
   \ref{f:mhifgas}) show that the average relation among irregular and 
   spiral galaxies is reasonably well matched with our recipe 
   (Eq. \ref{e:sfrscale}). The space between models with and without 
   infall events would evidently be occupied by events involving less 
   massive clouds, which presumably occur more often.

   However, the models have far less scatter in the gas fraction than is 
   evident from the observations. For masses less than about 
   $10^{8.5} \msun$ the infall event causes some variation of the 
   gas fraction. But for higher masses the region affected by the accreting
   gas cloud is rather small, and thus the overall gas fraction is determined
   by the global SFR, which we had assumed to be a unique function of the 
   galactic mass only. This assumption -- taken for reasons of simplicity -- 
   does not appear to be generally valid. We take this as evidence
   of a genuine variation of the star formation timescale between individual 
   galaxies, but we do not propose to model this here, because we wish to 
   avoid introducing further free parameters.

   In Fig. \ref{f:mhifgaso} one notes the presence of a few very low mass 
   galaxies with low observed gas fraction: UGC~12613 (type Im V, 
   $\lg(\mhi/\msun ) = 6.61$, $\lg(-\ln(f_{\rm gas})) = 0.35$),
   Pegasus  (T = 10, 6.82, 0.033), and UGC~11755 (BCD/E, 8.19, 0.46). 
   They could be interpreted in terms of our models as dwarf galaxies with
   a SFR higher than our recipe, having never undergone an accretion event, 
   and thus continuing to use up their original gaseous material.
   On the other hand, there are the more massive irregulars with high gas 
   fractions. The accretion event scenario might account for them, if gas 
   clouds of about $10^9$ to $10^{10}$ were involved. But they could also 
   merely have a SFR lower than the other galaxies.

 \subsection{Results: metallicity}

\begin{figure}
      \centering
      \includegraphics[angle=-90,width=10cm]{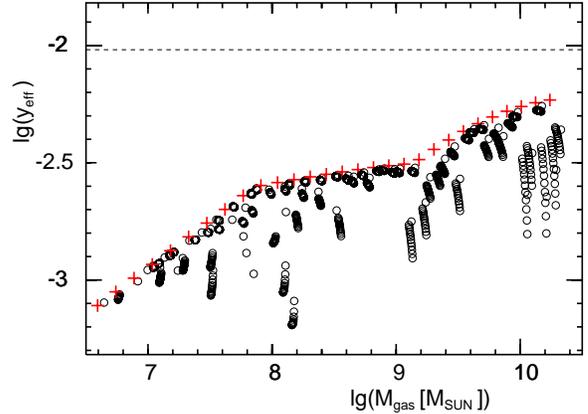} 
      \caption[]{Effective oxygen yields as a function of their instantaneous
                 \HI - mass of the models during the 2 Gyr after
                 the infall of a metal-poor gas cloud at 13 Gyr after start
                 of the evolution. The horizontal dashed line indicate the solar 
                 yield. The scales of this figure correspond exactly to those 
                 of Fig. \ref{f:mhiyeffo}.}
      \label{f:mhiyeff}  
    \end{figure}

\begin{figure}
      \centering
      \includegraphics[angle=-90,width=10cm]{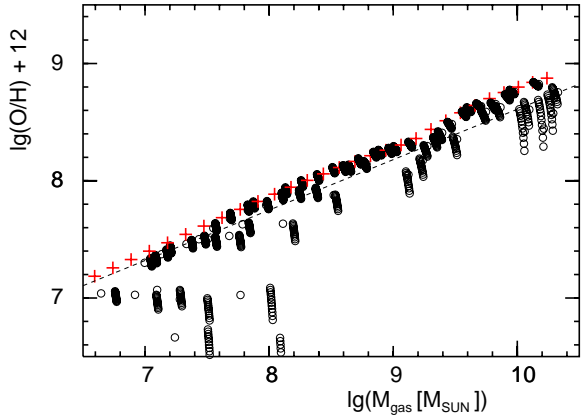} 
      \caption[]{The model oxygen abundances as a function of gas mass. The 
                 dashed line is the relation found by Garnett (2002).}
      \label{f:mhioh}  
    \end{figure}

   Our choice of the yield reduction factors (Eq. \ref{e:yieldfactor}) 
   gives a satisfactory match of the effective yields of the models 
   (Fig. \ref{f:mhiyeff}) with what is deduced from the observations 
   (Fig. \ref{f:mhiyeffo}). The observed mass-to-metallicity relation is
   also reasonably well reproduced by the models (Fig. \ref{f:mhioh}).

\begin{figure}
      \centering
      \includegraphics[angle=-90,width=10cm]{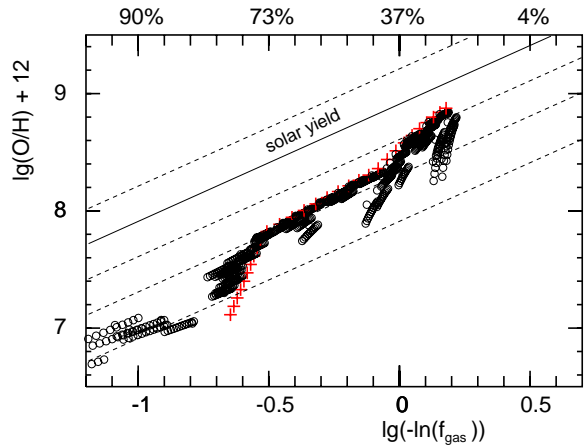} 
      \caption[]{Oxygen abundances as a function of the gas fraction for
                 our models.
                 The full line indicates the solar yield, the dashed lines
                 refer to twice, one half, one quarter, and one tenth
                 solar yield. For comparison with observations, see Fig.\
                 \ref{f:fgasoho}.}
      \label{f:fgasoh}  
    \end{figure}

    The oxygen abundances and gas fractions, in relation to the Simple Models
    are shown in Fig. \ref{f:fgasoh}. Comparison with Fig. \ref{f:fgasoho} confirms 
    that the choices for the dependencies of SFR and effective yields on galactic 
    mass are quite reasonable.
    It also emphasizes that the irregular galaxies of moderate 
    and high mass but large gas fractions (see Fig. \ref{f:fgasoho}) 
    are not well represented. While
    the two assumed relations could not model this, the episodic infall also 
    cannot account for the presence of these galaxies. The low surface 
    brightness ``quiescent" irregular galaxy DDO 154, with its high gas
    fraction and high yield but otherwise with average properties
    (cf. Kennicutt \& Skillman 2001) is a particularily
    striking example, as shown in Fig. \ref{f:fgasoho}. 

   Our recipes for star formation and yield were chosen so that the resulting 
   models give a reasonable match of the dependences of metallicity, gas 
   fraction, and effective yield with the gas mass. The oxygen abundances for 
   massive galaxies thus are slightly higher than Garnett's (2002) mean
   relation, just as in the observations (Fig. \ref{f:mhioho}). Among the 
   dwarf galaxies, the models seem to indicate the presence of many objects 
   with very low metallicities. However, one must keep in mind that here the 
   accretion of small clouds will dominate: events with clouds less massive 
   than the minimum of $10^6 \msun$ we show in the figures would fill up the
   space in the figures between the locus of the crosses and the model 
   tracks with infall. As these events can be expected to be more frequent than 
   collisions with larger clouds, the emphasis in the diagram would be 
   towards the less perturbed systems.

 \subsection{Results: the N/O-O/H Diagram}

\begin{figure}
      \centering
      \includegraphics[angle=-90,width=10cm]{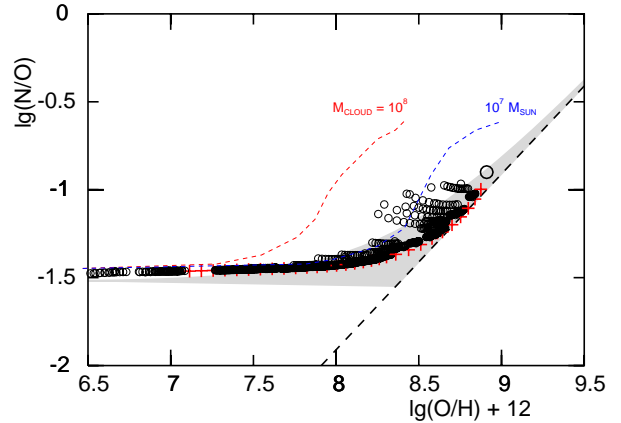} 
      \caption[]{The abundance ratio N/O as a function of oxygen abundance
                 for the models. The symbols are as in Fig. \ref{f:mhifgas}.
                 The crosses which denote models without infall are only
                 barely visible and form the lower limit for the plotted 
                 circles. The short-dashed curves envelope the maximum 
                 loop excursions for cloud masses of $10^7$ and $10^8 \msun$.  
                 The dashed straight line indicates the secondary nitrogen 
                 production track as in Fig. \ref{f:hnooho}.}
      \label{f:nooh}  
    \end{figure}

   In the diagram of nitrogen-to-oxygen abundance ratio versus
   oxygen abundance (Fig. \ref{f:nooh}), excursions from the closed-box
   relation (indicated by the chain of plus-signs) are rather large but 
   limited to a reduction in O/H in the horizontal part i.e. involving less 
   massive galaxies. In the ascending part there are also quite strong 
   effects due to more massive and hence metal-rich galaxies, which may 
   populate the region above the horizontal part.

\begin{figure}
      \centering
      \includegraphics[angle=-90,width=10cm]{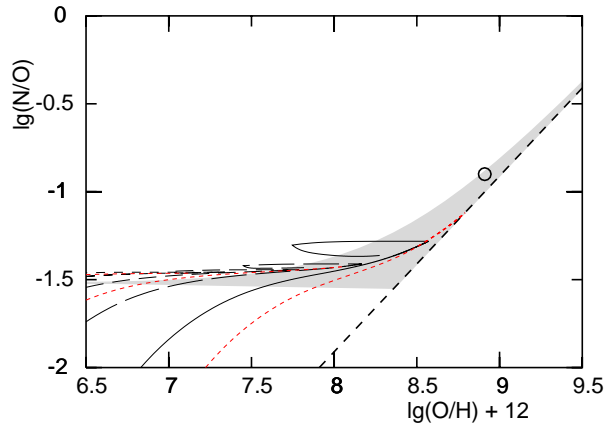} 
      \caption[]{Complete evolutionary tracks in the N/O-O/H diagram: short 
          dashed curves denote two models without infall events (galactic 
          masses $10^{8.5}$ and $10^{10.5} \msun$); the others are models 
          with a $10^8 \msun$ cloud falling on to a galaxy with masses of 
          $10^{10} \msun$ (full line), and $10^9$, $10^8$, 
          and $10^7$ (with progressively shorter dashes).}
      \label{f:nooh1}  
    \end{figure}

   If one looks at the evolutionary tracks in this diagram (Fig.
   \ref{f:nooh1}), one notes that -- by construction -- low mass galaxies 
   due to their longer star formation timescale evolve only until the 
   horizontal part of the observational diagram, while more massive ones 
   can reach up to the ascending branch. For instance the model of 
   $10^{10} \msun$ evolved (up to the present time) to a gas fraction of 
   about 10\% and an oxygen abundance somewhat larger than the sun, when the
   infall occurs. That a rather large loop is obtained has three reasons: 
   firstly, a $10^8 \msun$ cloud affects only 0.002 of the galaxy's gas mass
   i.e. about $2\,10^7 \msun$ which yields a reduction of the oxygen 
   abundance by nearly 0.8 dex. Secondly, the star formation timescale is 
   rather short -- 1.06 Gyr -- and thus during the 100 Myr of the infall an 
   appreciable quantity of oxygen is produced, and the N/O ratio can also be 
   reduced (cf. Fig. \ref{f:noldur}). Thirdly, the high star formation leaves
   a smaller gas mass in the galaxy at the start of infall. Note that during 
   the infall episode the loop is passed through quickly, thus there is 
   little chance to detect a galaxy in such a state. It is more likely to be 
   found after the event when it evolves towards 
   the closed-box relation. The enhanced density of symbols on these upward 
   curved tracks makes them easily identified in Fig. \ref{f:nooh}.
   Since the infall onto a less massive galaxy (e.g. $10^7 \msun$) affects
   the entire galaxy, the effects on abundances are also quite substantial, 
   though not too evident in the diagram: the reduction of oxygen abundance 
   is about 1 dex. But the evolutionary track after the event is at nearly 
   constant N/O ratio, due to the slower SFR. This gives sufficient time for 
   a stellar generation to eject all the delayed primary nitrogen from 
   intermediate mass stars along with the undelayed oxygen. 

   As is evident from the tracks before the infall event, their slope becomes 
   steeper in more massive galaxies. This is because the higher star formation
   rate makes the effect of the time-delay of the nitrogen production in 
   intermediate mass stars more pronounced. Thus, it is possible that the 
   loop of a more massive galaxy may descent somewhat below the tracks of 
   less massive galaxies.

\section{Summary and discussion}

    Our model calculations demonstrate that it is possible to obtain a
    significantly broadened N/O-O/H relation by the galaxies evolving 
    towards the closed-box relation after having undergone an episode of 
    strong infall. 
  
    One prediction of our scenario is that galaxies at high N/O ratio would 
    tend to be more massive, as they originate from galaxies which would have
    evolved into the ascending branch, and the infall would have preserved
    the enhanced N/O ratio. Also, we expect that in the horizontal part and 
    in the lower ascending part there is a mixture of less massive galaxies
    without 
    an infall event which experience a long evolution due to slow star
    formation and of more massive ones having undergone an infall episode. 
    Comparison with Fig. \ref{f:nooho} shows a substantial amount of mixture, 
    but the observational uncertainties in the abundances and the lack of 
    \HI mass determinations do not permit to clearly confirm or disprove the 
    expected trends.

  \subsection{The N/O ratio}

\begin{figure}
      \centering
      \includegraphics[angle=-90,width=10cm]{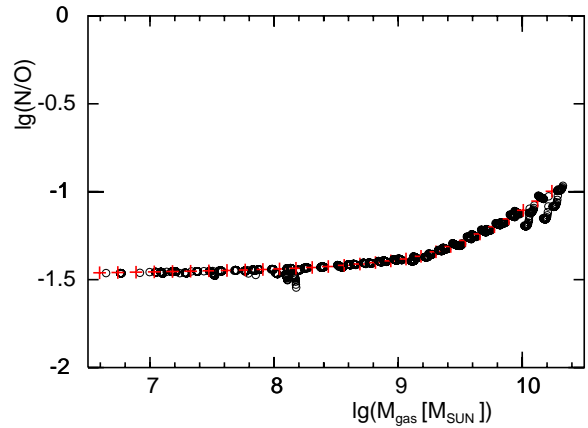} 
      \caption[]{The N/O abundance ratio as a function of gas mass,
               as computed from the models.}
      \label{f:mhino}  
    \end{figure}

    The computed N/O ratios as a function of galactic \HI mass,
    presented in Fig. \ref{f:mhino}, are nearly constant for small masses,
    since the loops due to infall events take place in the flat part of
    the N/O-O/H diagram (cf. Fig. \ref{f:hekdur}). Only for high mass models
    does the ratio increase by about 0.5 dex. This is because  the
    massive objects would cover the ascending branch. Here, infall 
    episodes may lower the N/O ratio, but only by about 0.2 dex.
    
\begin{figure}
      \centering
      \includegraphics[angle=-90,width=10cm]{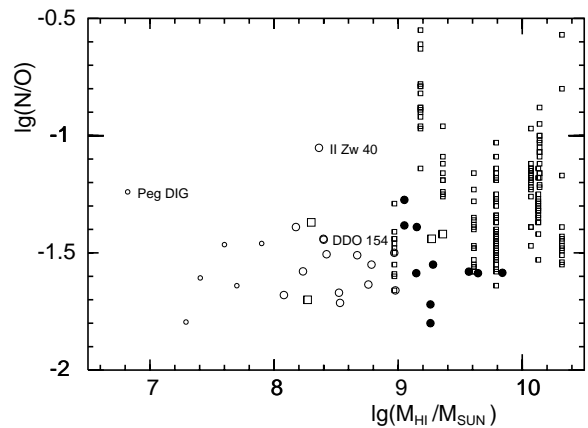} 
      \caption[]{The observed N/O abundance ratio as a function of 
                  galactic \HI mass. The data from spiral galaxies (small 
                  squares) pertain to individual \HII regions, while the 
                  \HI mass of the parent galaxy is used. The other symbols are 
                  the same as in Fig. \ref{f:mhioho}.}
      \label{f:mhinoo}  
    \end{figure}

    Among the observations (Fig. \ref{f:mhinoo}) we note that apart 
    from II Zw 40 and the Pegasus dwarf, the irregular galaxies show
    no clear trend with \HI mass, but seem to scatter around a constant 
    value. In this figure none of the objects marked with crosses in 
    Fig. \ref{f:nooho} is present. The \HII region data from spiral galaxies
    show the range of N/O ratios met in these objects. On average, the 
    values are slightly higher than among the irregular galaxies.
 
    For II Zw 40, Kobulnicky \& Skillman (1996) give $\lg({\rm N/O}) = -1.05$
    while Izotov \& Thuan (1999) give $-1.53$ which is more like the other
    objects. The value for the Pegasus dwarf of $-1.24$ is found by
    Skillman et al. (1997) to be one sigma above the values of the other
    irregular galaxies. For this exceptional object, the authors discuss
    several possible explanations, including a different IMF and time delayed
    nitrogen production. This would remain valid in the context of episodic 
    infall. We shall not consider these two cases. We also note that our 
    models with their single dependence of parameters on galactic mass 
    cannot account for the appreciable scatter in the diagram. 

  \subsection{Models with additional starbursts}

\begin{figure}
      \centering
      \includegraphics[angle=-90,width=10cm]{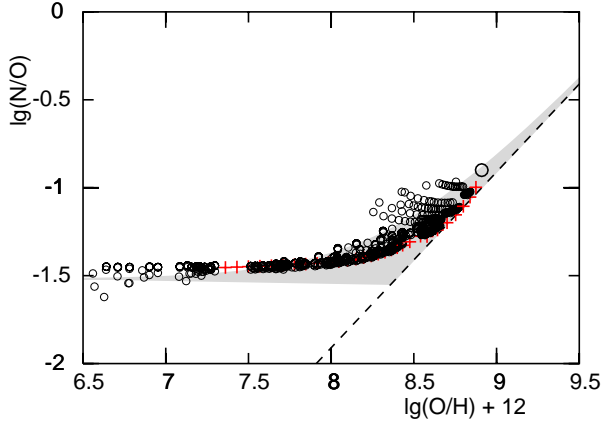} 
      \caption[]{The N/O-O/H diagram for models with additional 
                      genuine star bursts.}
      \label{f:snooh}  
    \end{figure}

\begin{figure}
      \centering
      \includegraphics[angle=-90,width=10cm]{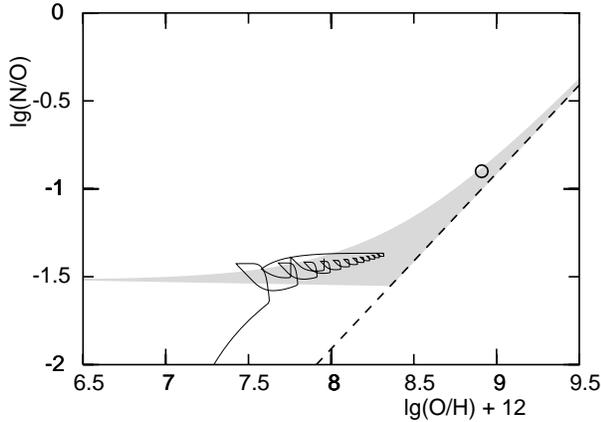} 
      \caption[]{Evolutionary tracks in the N/O-O/H diagram for a model with
              $10^9 \msun $ mass and with genuine star bursts.}
      \label{f:snooh1}  
    \end{figure}

   To explore the additional effects due to the presence of genuine
   starbursts, we compute the same series of models with a bursty
   star formation history. We apply the following ad-hoc recipe:
   for all galaxies less than $10^{10} \msun$ the star formation
   rate is 10 times the continuous rate used before, for a duration
   of 200 Myr beginning every full Gyr, but during the 800 Myrs 
   in between the rate is one hundred times less the continuous rate.
   This increases the scatter in the N/O-O/H diagram (Fig. \ref{f:snooh})
   somewhat. One notes that the closed-box relation is no longer as
   strictly obeyed as a limit; as expected, several points lie below the 
   relation.
   Likewise, in all the other diagrams the amount of scatter among the dwarf 
   galaxy models is further increased, although this does not significantly 
   change the relations seen in Figs. \ref{f:mhifgas} to \ref{f:mhioh}.
   The evolutionary tracks (Fig. \ref{f:snooh1}) before the infall event show 
   the typical Garnett zig-zag curves, although somewhat modified because
   of the continuous infall. The accretion event gives a large loop, with
   a variation of about 0.1 dex in the N/O ratio. 

\section{Conclusions}

   We examine the chemical evolution following the rapid infall of metal-poor
   gas into a galaxy. First, we consider models with a simplified
   nucleosynthesis recipe and explore the influence of the various parameters,
   and under which circumstances one achieves large deviations from the
   closed-box Simple Model. We find that
   \begin{itemize}
     \item excursions in the  N/O-O/H diagram are in the form of loops
           but only towards the `upper-left' side of the closed-box
           relation, that is towards smaller oxygen abundances and 
           larger N/O ratios.
     \item during the infall, the oxygen abundance is decreased due
           to dilution of the galactic gas, followed after the infall by
           the evolution towards the closed-box relation.
     \item in order to have large excursions, the mass of the infall gas
           must be much larger than the mass of the gas present in the galaxy 
           (or in the star-forming region of large galaxies).
     \item also, the infall rate must be greater than the SFR.
     \item in order to form large loops, the duration of the event must be
           of the order of the star formation timescale.
  \end{itemize}
   These results, which may also be obtained by analytical solutions, are
   also found from numerical models with a detailed nucleosynthesis
   recipe and taking into account the finite stellar lifetimes.

   We then examine the possible effects of HVCs colliding with gas-rich 
   galaxies on the chemical evolution of the galaxy. From the properties 
   attributed to HVCs we estimate that  
   \begin{itemize}
      \item a passage would occur for 10 to 100 Myr which is suitably
            short compared to the star formation,
      \item the gas masses of $10^6$ to $10^9 \msun$ would be sufficient
            to cause a large influence on dwarf galaxies,
      \item in massive (and large) galaxies the HVCs size of 1 to 10 kpc 
           would restrict any observable effects to a partial volume of the
           galaxy.
   \end{itemize}
   This implies that such a collision could be expected to leave observable
   marks in the chemical properties of the galaxies, different from what
   a closed-box evolution predicts.
   Based on the experiences of the simplified models -- but confirmed with
   the detailed models, the above characteristics also imply that 
   \begin{itemize}
      \item full loops in the N/O-O/H diagram are not realized, because the
            infall duration is short,
      \item large loops or excursions will be rare, since they require the
            encounter with a massive cloud component which 
            most probably is a rare species,
      \item because of the short infall duration, galaxies are most likely
            to be observed during the slow evolution back towards the
            closed-box relation. If infall triggers enhanced star formation,
            the galaxies might be very bright at that early stage.
  \end{itemize}

   To model the global properties of galaxies of various masses, we
   assumed relations of star formation timescale and yield on
   galactic mass which we adjusted to reproduce the observed 
   average relations of gas fraction, oxygen abundance, and effective
   yield with \HI mass. Evidently, our models cannot account for the
   appreciable dispersion of the observed galactic properties, because
   e.g. infall happens in reality during all stages of evolution.
   Following the chemical evolution for the 2 Gyrs after a single encounter 
   with a HVC, we find that among galaxies less massive than about
   $10^8 \msun$ the deviations from models without infall events
   are substantial enough to provide some contribution to the observed
   scatter of galactic properties.
    
   In the N/O-O/H diagram, the galaxies are mainly caught on the slow 
   way back to the closed-box relation. Bright starburst galaxies
   might be observable at earlier stages, such as near the largest 
   decrease in oxygen abundance. All this adds a quite significant
   dispersion towards the `upper-left' side. Thus the dispersion of
   irregular galaxies towards N/O ratios higher than the plateau at
   $\lg({\rm N/O}) = -1.5$ might well be objects that had evolved until
   higher metallicities than they now have, but which suffered an 
   infall episode with a HVC.

   On the other hand, we note that the deviations remain somewhat 
   limited; galaxies are likely to be observed in the vicinity of the 
   closed-box relation. Thus, the N/O-O/H plot retains its 
   sensitivity to the way the two elements are produced and incorporated 
   in the interstellar gas. The diagram remains a very useful tool to 
   infer details about stellar nucleosynthesis.
%__________________________________________________________________

  \begin{acknowledgements}
    The authors are grateful to Andreas Rieschick for stimulating discussions. 
    J.K.\ acknowledges support by the University of Vienna for his visit
    and the hospitality of the Vienna Institute of Astronomy where the 
    studies were completed.
  \end{acknowledgements}

%__________________________________________________________________


\begin{thebibliography}{}

\bibitem[]{}
   Blitz L., Spergel D.N., Teuben P.J., Hartmann D., Burton W.B.,
      1999, ApJ 514, 818
\bibitem[]{}
   Braun R., Burton W.B., 1999, A\&A 341, 437
\bibitem[]{}
   Braun R., Burton W.B., 2000, A\&A 354, 853
\bibitem[]{}
   Broeils A.H., van Woerden H., 1994, A\&AS 107, 129
\bibitem[]{}
   Broeils A.H., Rhee M.-H., 1997, A\&A 324, 877
%\bibitem[]{}
%   Burton W.B., Braun R., Chengalur J.N., 2000, A\&A 369, 616
\bibitem[]{}
   Carignan C., Purton C., 1998, ApJ 506, 125
\bibitem[]{}
   de Heij V., Braun R., Burton W.B., 2002, A\&A 392, 417
\bibitem[]{}
   Duc P.-A., Papaderos P., Balkowski C., Cayatte V., 
     Thuan T.X., van Driel W., 1999, A\&AS 136, 539
\bibitem[]{}
   Duc P.-A., Cayatte V., Balkowski C., Thuan T.X., Papaderos P., 
     van Driel W., 2001,  A\&A 369, 763
\bibitem[]{}
   Edmunds M.G., Pagel B.E.J., 1978, MNRAS 185, 78P
\bibitem[]{}
   Garnett D.R., 1990, ApJ 360, 142
\bibitem[]{}
   Garnett D.R., Skillman E.D., Dufour R.J., et al., 1995, 
     ApJ 443, 64
\bibitem[]{}
   Garnett D.R., 2002, ApJ 581, 1019
\bibitem[]{}
   Henry R.B.C., Worthey G., 1999, PASP 111, 919
\bibitem[]{}
   Henry R.B.C., Edmunds M.G., K\"oppen J., 2000, ApJ 541, 660
\bibitem[]{}
   Izotov Y.I., Thuan T.X., 1999, ApJ 511, 639
\bibitem[]{}
   Kennicutt R.C., Skillman E.D., 2001, ApJ 121, 1461
\bibitem[]{}
   Kilborn V.A., Stavely-Smith L., Marquarding M., et al.,
       2000, AJ 120, 1342
\bibitem[]{}
   K\"oppen J., Edmunds M.G., 1999, MNRAS 306, 317
\bibitem[]{}
   Kobulnicky H.A., Skillman E.D., 1996, ApJ 471, 211
\bibitem[]{}
   Maeder A., 1992, A\&A  264, 105
\bibitem[]{}
   Mateo M., 1998, ARAA 36, 435
\bibitem[]{}
   Matteucci F., 1986, PASP 98, 973
\bibitem[]{}
   Pisano D.J., Kobulnicky H.A., Guzman R.,
      Gallego J., Bershady M.A.,  2001, ApJ 122, 1194 
\bibitem[]{}
   Salzer J.J., Rosenberg J.L., Weisstein E.W., Mazzarella J.M., 
   Bothun G.D., 2002, AJ 124, 191
\bibitem[]{} 
   Skillman E.D., Bomans D.J., Kobulnicky H.A., 1997, ApJ 474, 205
\bibitem[]{}
   van den Hoek, Groenewegen M.A.T., 1997, A\&AS 123, 305
\bibitem[]{}
   van Zee L., 2001, AJ 121, 2003
\bibitem[]{}
   van Zee L., Haynes M.P., Salzer J.J.,  1997a, AJ 114, 2497
\bibitem[]{}
   van Zee L., Haynes M.P., Salzer J.J., Broeils A.H., 
      1997b, AJ 113, 1618
\bibitem[]{}
   van Zee L., Salzer J.J., Haynes M.P., 1998a,
      ApJ 497, L1
\bibitem[]{}
   van Zee L., Salzer J.J., Haynes M.P., O'Donoghue A.A.,
      Balonek T.J., 1998b, AJ 116, 2805
\bibitem[]{}
   van Zee L., Skillman E.D., Salzer J.J.,  1998c, AJ 116, 1867
\bibitem[]{}
   van Zee L., Salzer J.J., Skillman E.D., 2001, AJ 122, 121
\bibitem[]{}
   Vila-Costas M.B., Edmunds M.G., 1993, MNRAS 269, 199
\bibitem[]{}
   Young L.M., Lo K.Y., 1996, ApJ 462, 203
\bibitem[]{}
   Young L.M., van Zee L., Lo K.Y., Dohm-Palmer R.C., Beierle M.E.,
      2003, ApJ 592, 111
\end{thebibliography}
\end{document}